\documentclass[11pt,twoside]{article}

\usepackage{asp2004}
\usepackage{epsf}
\usepackage{psfig}
\usepackage{lscape}

\begin{document}
\title{Identifying and Characterizing New Nearby White Dwarfs}   

\author{John P. Subasavage$^1$, 
        Todd J. Henry$^1$,
        Pierre Bergeron$^2$,
        Patrick Dufour$^2$,
	Nigel C. Hambly$^3$,
        Thomas D. Beaulieu$^1$}           

\affil{$^1$ Department of Physics and Astronomy, Georgia State
University, 1 Park Place, Suite 700, Atlanta, GA 30302-4106 \\
$^2$ D\'{e}partment de Physique, Universit\'{e} de Montr\'{e}al, CP 6128,
Succursale Centre-Ville, Montreal, QC H3C 3J7, Canada \\
$^3$ Institute for Astronomy, University of Edinburgh Royal Observatory,
Blackford Hill, Edinburgh EH9 3HJ, Scotland, UK}

\begin{abstract} 

How confident are we that all of the nearest white dwarfs (WDs) have
been identified?  In an effort to answer this question, we have begun
an initiative to identify and characterize new nearby WDs,
particularly in the southern hemisphere.  We estimate physical
parameters for new WDs using medium resolution (R $\sim$1000) optical
spectroscopy, and distances using optical photometry combined with
2MASS near-infrared photometry.  For objects within 25 pc (Catalogue
of Nearby Stars, and NStars Database horizons), we determine a
trigonometric parallax via CTIOPI (Cerro Tololo Inter-American
Observatory Parallax Investigation).  Of the 37 new WD systems
discovered so far, fourteen are likely within 25 pc, a volume that
contains 107 WDs with trigonometric parallaxes.  Interesting objects
include two that are likely double degenerates including one with a
magnetic component, one that is a cool (T$_{eff}$ $\sim$5000 K) likely
mixed atmosphere WD with deficient flux at near-infrared wavelengths,
and two that are metal-rich.  Observations are underway via the Hubble
Space Telescope to resolve four potential double degenerates (the new
magnetic WD and three other previously known WDs) for dynamical mass
determinations.  All ground-based observations are obtained as part of
the SMARTS (Small and Moderate Aperture Research Telescope System)
Consortium at CTIO.

\end{abstract}


\section{Introduction}   

White dwarfs (WDs) are the end results for all stars less than $\sim$
6 - 8 M$_{\sun}$; therefore they are relatively numerous.  WDs have
been used as photometric and spectroscopic calibrators, as proxies for
galactic population ages, and as constraints for stellar evolution
theory.  Intrinsic faintness makes WDs hard to study, yet a complete
sample of the nearest WDs to the Sun (i.e. the brightest
representatives of this class of stars) is essential to understand the
precursor population.

Our goal is to identify previously unknown WDs in the solar
neighborhood, to estimate distances, and to measure trigonometric
parallaxes and confirm proximity.  In addition, we estimate distances
for previously known WDs without trigonometric parallaxes and obtain
parallax determinations for objects presumed to be within 25 pc
(Catalogue of Nearby Stars and NStars Database horizons).  All of the
37 newly identified WD systems reported here are brighter than $V$ $=$
17.0 and most are in the southern hemisphere, a region sampled more
poorly than in the northern hemisphere.  Although this new sample is
relatively small, we find several interesting objects, exemplifying
the vast diversity seen in much larger WD samples, some of whose
characteristics are not well understood within current theoretical
framework.

\section{Candidate Selection}

Proper motion is the most definitive means to tease out WDs from the
far more populous and distant main sequence stars that look identical
photometrically.  In collaboration with Nigel Hambly at the Royal
Observatory at Edinburgh, Scotland, we have completed a survey of the
southern sky for high proper motion objects using astrometric and
photometric data from the SuperCOSMOS Sky Survey (SSS)
\citep{2004AJ....128..437H, 2004AJ....128.2460H, 2005AJ....129..413S,
2005AJ....130.1658S}.  The SuperCOSMOS-RECONS (SCR) proper motion
survey, in addition to discovering nearly 300 new high proper motion
objects in the southern sky with 10.0 $<$ $R_{59F}$ $\le$ 16.5 and
0.4$\arcsec$/yr $\le$ $\mu$ $\le$ 10.0$\arcsec$/yr, has recovered
thousands of known proper motion objects.  Because of the quality of
the photometric calibration of the SSS plates, the data for these
known objects are superior to the magnitude data of the original
discovery survey (i.e. \citealt*{luyten}), enabling us to better
identify WD candidates.  We compliment the SSS photometry with
near-infrared 2MASS $JHK_s$ photometry.  Utilizing both optical plate
and near-infrared photometry and the SSS astrometry, we calculate the
($R_{59F} - J$) color and the reduced proper motion (RPM) as follows:

\vspace{-5pt}
\begin{equation}
H_{R_{59F}} = R_{59F} + 5 + 5 \log(\mu)
\end{equation}

\begin{figure}[!ht]
\plotfiddle{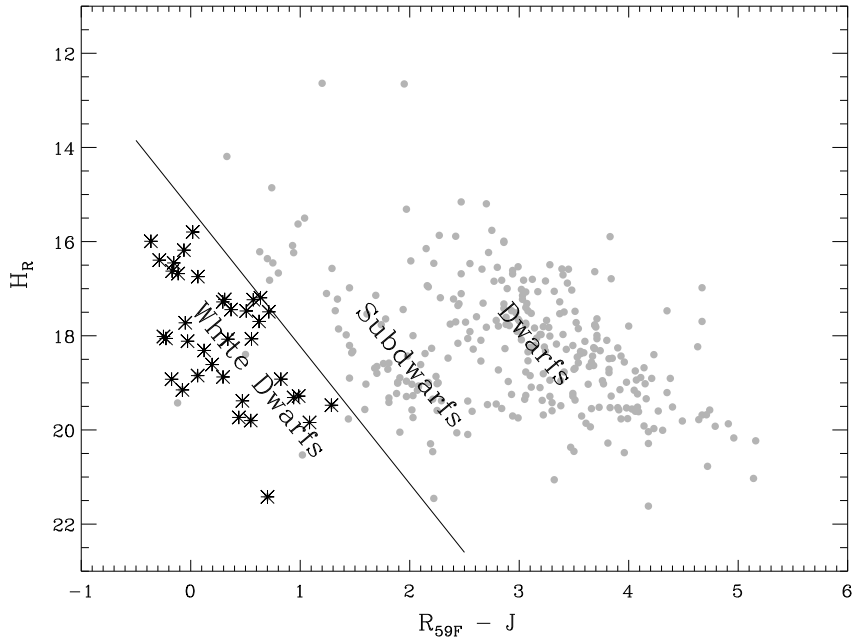}{5in}{90}{50}{50}{195pt}{100pt}
\vspace{-120pt}
\caption{Reduced proper motion diagram that illustrates the locations
of the three luminosity classes.  Small filled circles represent the
SCR proper motion sample.  Asterisks represent the 37 new WD systems
reported here.  The diagonal line is an arbitrary delineator above
which subdwarfs are generally found and below which WDs are generally
found.}
\label{red}
\end{figure}

RPM correlates proper motion ($\mu$) with proximity, an association
that is certainly not always valid but is sufficient to delineate the
luminosity classes.  Figure \ref{red} illustrates this technique.  The
filled circles are the SCR proper motion discoveries plotted to show
the three distinct groupings.  The asterisks are the 37 new WD
discoveries presented in this work.  The diagonal line separating the
WDs from the subdwarfs was drawn arbitrarily to serve as an
approximate dividion between the two luminosity classes.  There are
three objects well within the WD region that are not marked with
asterisks because we have not yet obtained spectra, although they are
most likely WDs.  We have confirmed a fourth object just below the
line at ($R_{59F} - J$) $=$ 1.4 (not marked with an asterisk) to be a
subdwarf contaminant.

\section{Spectroscopy}

After selection of WD candidates via the RPM diagram, spectra were
obtained at the CTIO 1.5m telescope operated by the SMARTS (Small and
Moderate Aperture Research Telescope System) Consortium.  Observations
were carried out during several observing runs between July 2003 and
May 2006.  The Ritchey-Chr\'{e}tien Spectrograph and Loral
1200$\times$800 CCD detector were used with grating \#09, providing
8.6 \AA~resolution and wavelength coverage from 3500 \AA~to 6800 \AA.
Bias subtraction, dome / sky flat-fielding, and extraction of spectra
were performed using standard IRAF packages.  Integration times were
typically several minutes up to 30 minutes.  Two spectra were taken in
series for each object to allow cosmic ray rejection.

Spectra of the 37 new WD systems indicate that 30 are DA, four are DC,
two are DZ, and one is DQ.  


\section{Photometry}

Using a subsample of WDs from the \citet{2001ApJS..133..413B} sample
of WDs with $BVRIJHK$ photometry and trigonometric parallaxes, we
created a suite of color relations linked to $M_V$.  These relations
can be used to estimate distances of WDs for which no trigonometric
parallax is available.  We obtained multi-epoch $V_JR_{KC}I_{KC}$
photometry at the CTIO 0.9m telescope and extracted 2MASS $JHK_s$ for
the 37 new WD systems and other known WD systems for which no
trigonometric parallax exists.  The photometry allowed us to (1)
estimate a distance using the suite of relations, and (2) reproduce
the spectral energy distribution (SED) and obtain an effective
temperature (assuming either H or He atmosphere, which can be
constrained by the spectrum).  Table \ref{photdist} displays the
number of new and known WDs presumed to be within 25 pc.  The
trigonometric parallax is needed to constrain log $g$ and other
parameters dependent on log $g$ (i.e. mass, age, and luminosity).
Because these objects do not yet have trigonometric parallaxes, a log
$g$ of 8 is assumed -- an entirely valid assumption when compared to a
large WD sample for which log $g$ is constrained via trigonometric
parallax (i.e. \citealt*{2001ApJS..133..413B}).  However, other
dependent parameters may not be well constrained and are disregarded
in this analysis.

\begin{table}[!ht]
\caption{Photometric Distance Estimates for New and Known WDs Without
Trigonometric Parallaxes
\label{photdist}}
{
\begin{tabular}{lccc}
\noalign{\smallskip}
\tableline
\noalign{\smallskip}
Sample  &~~ D $\le$ 10 pc ~~ &~~ 10 pc $<$ D $\le$ 25 pc~~ &~~  D $>$ 25 pc~~ \\
\noalign{\smallskip}
\tableline
\noalign{\smallskip}

New        &  $0$  & $14$ &  $23$  \\
Known      &  $2$  & $15$ &  $ 0$  \\
\noalign{\smallskip}
\tableline
\noalign{\smallskip}

Total  &  $2$  & $29$ &  $23$  \\

\noalign{\smallskip}
\tableline

\end{tabular}
}
\end{table}

\vspace{-10pt}
\section{Astrometry}

In order to be certain that a WD is indeed nearby, a trigonometric
parallax is needed.  For objects estimated to be within 25 pc, we
determine a trigonometric parallax via the 0.9m Cerro Tololo
Inter-American Observatory Parallax Investigation (CTIOPI) program
\citep{2005AJ....129.1954J}.  To date, there are 107 WDs with
trigonometric parallaxes within 25 pc.  One of our goals is to
increase this sample by 25\% or 27 new systems.  Currently there are
44 WD systems on the parallax program and of those, 23 have
preliminary parallaxes placing them within 25 pc.  We are confident
that our goal will be realized.

\section{Interesting Objects}

Once all of the previously mentioned data are collected, we are able
to characterize these objects precisely.  In some cases, interesting
attributes present themselves when all the data are available.  Here
we highlight a few of the interesting new objects.

\subsection{WD 0622-329}

\begin{figure}[!ht]
\plottwo{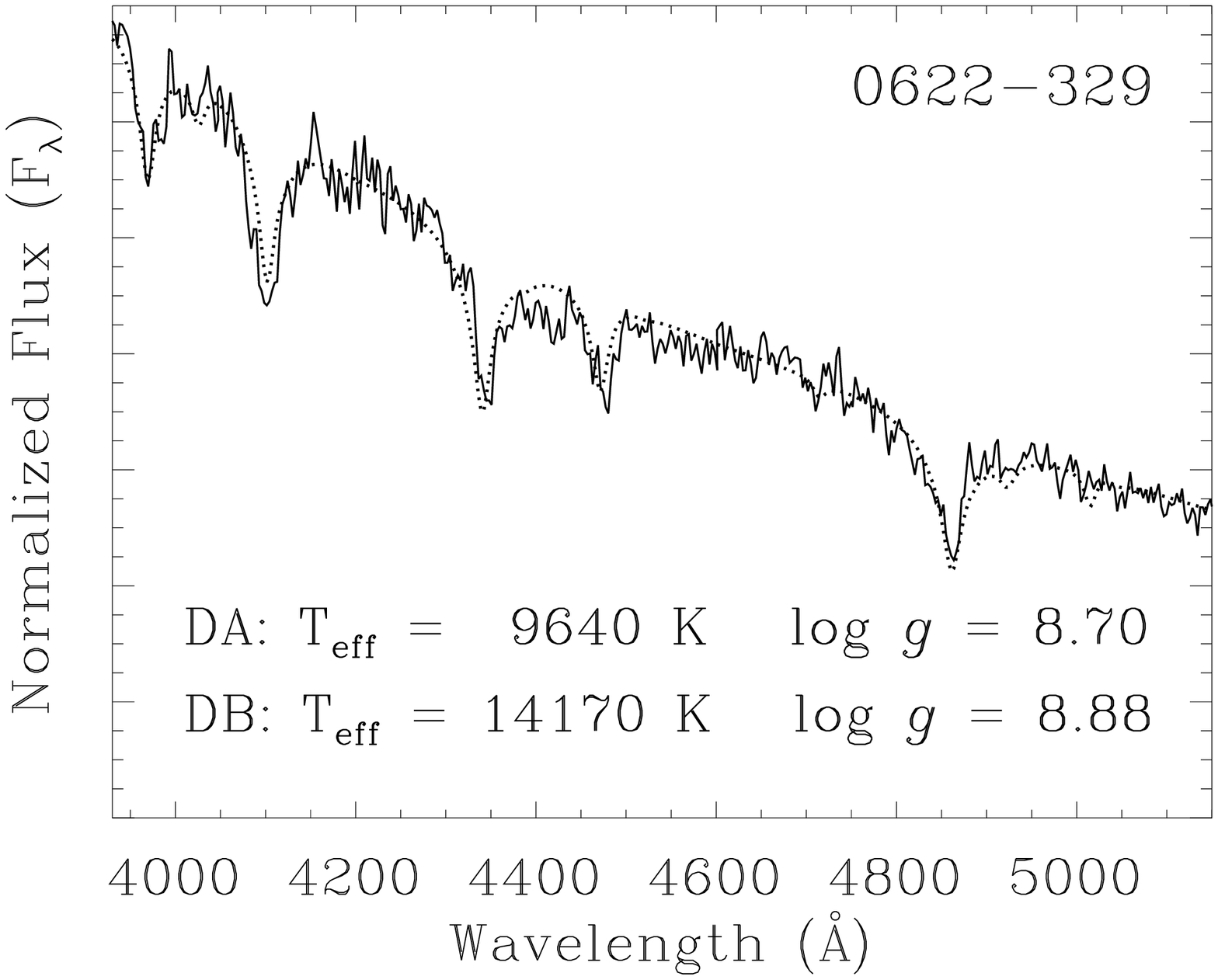}{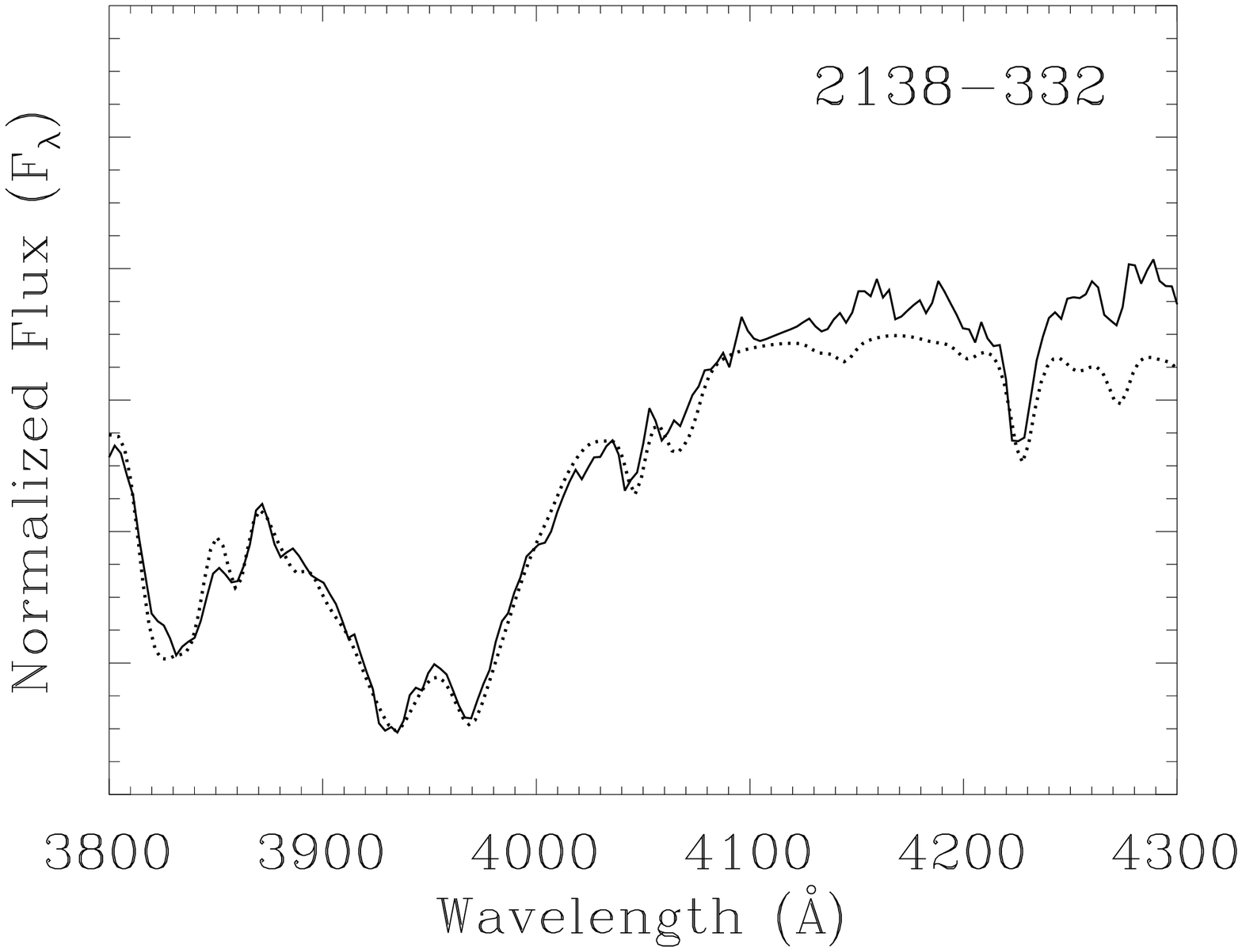}
\caption{(\emph{a}) Spectral fit for WD 0622-329 assuming an
unresolved double degenerate with one component being a DA and the
other being a DB.  (\emph{b}) Spectral fit for new DZ white dwarf WD
2138-332.}
\label{dadbdz}
\end{figure}

This WD has both H and He lines in its spectrum.  When we attempt to
fit the spectrum, we get a $T_{eff}$ $\sim$43,000 K and the fit is
moderately good except that it predicts He {\scriptsize II} absorption
at 4686 \AA, which is certainly not seen in the spectrum.  If we use
the photometry and perform a SED fit, we obtain $T_{eff}$ $\sim$10,500
K.  We then tried to reproduce the spectrum assuming the WD is an
unresolved double degenerate with one component's atmosphere being H
and the other being He.  The fit (dashed line) is shown in Figure
\ref{dadbdz}\emph{a} and a convolution of the two temperatures is
consistent with the SED fit.

\subsection{WD 2138-332}

One of only two DZ WDs from the new sample, this object exhibits
strong absorption due to Ca and Mg.  Theoretical models have only
recently provided accurate treatment for these objects such that
spectral features are reproduced reliably.  Figure
\ref{dadbdz}\emph{b} exemplifies this point.  The solid line
represents the observational data while the dashed line represents the
model.  This object is well characterized as a DZ at $T_{eff}$ $=$
7188 K and a log N(Ca)/N(He) $=$ -8.64.

\subsection{WD 2008-600}

\begin{figure}[!ht]
\plottwo{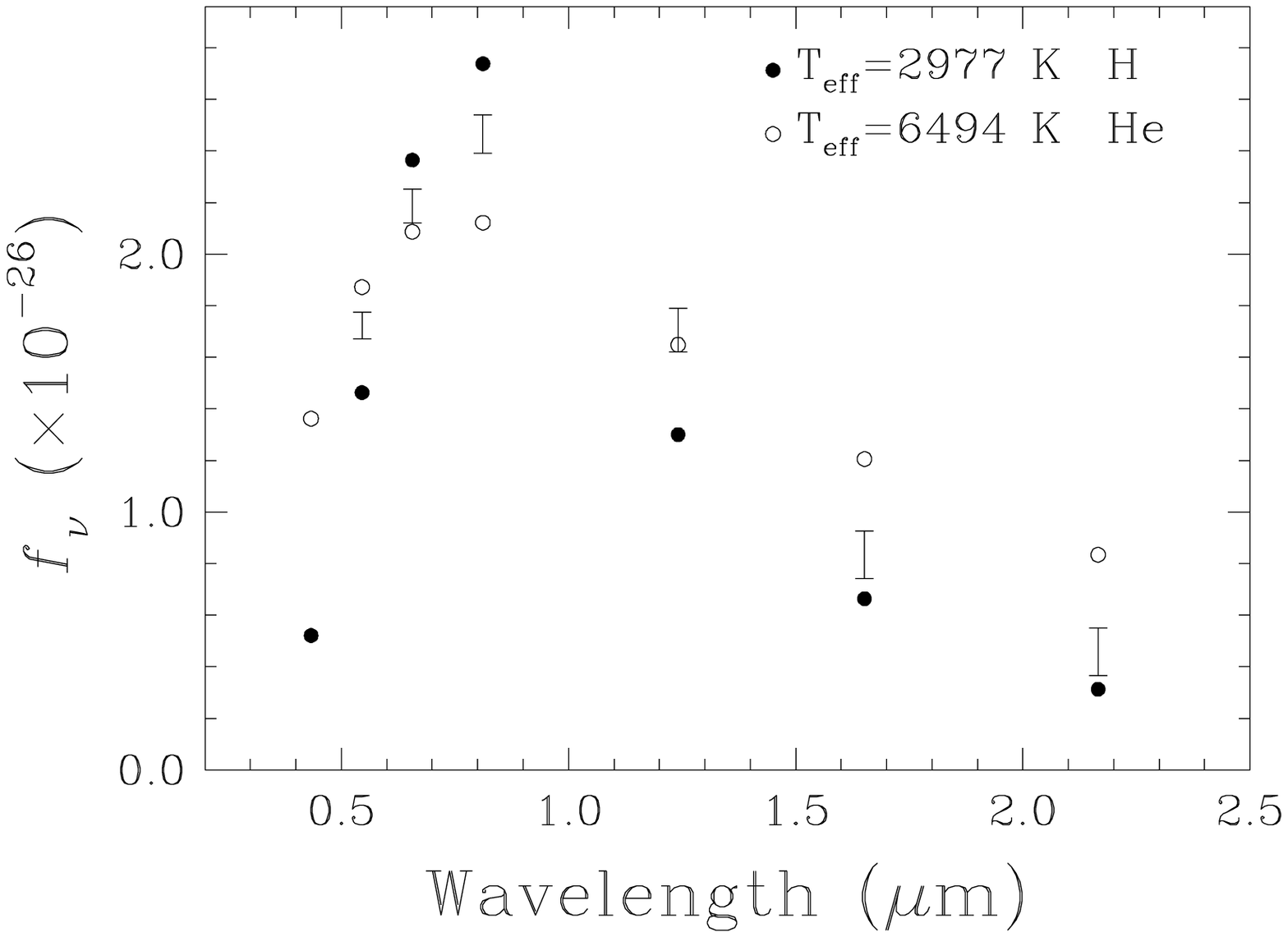}{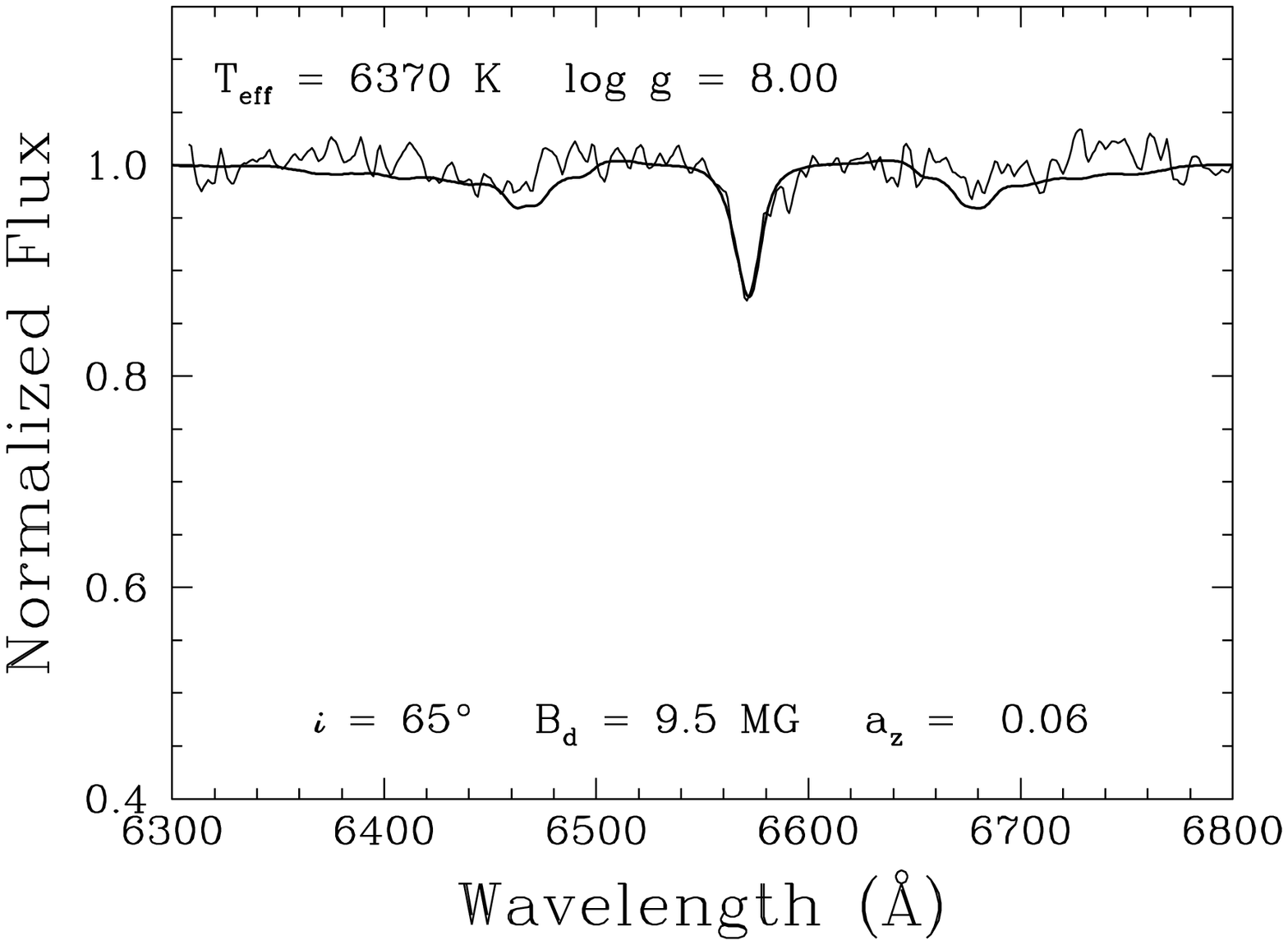}
\caption{(\emph{a}) SED fit for WD 2008-600.  The error bars represent
the photometry values.  (\emph{b}) Spectral fit for WD 0121-429 with
the heavy line representing the fit and the lighter line representing
the observational data.}
\label{dcdah}
\end{figure}

This object exhibits no features in its spectrum.  Using the
photometry for the SED fit, one obtains vastly different $T_{eff}$
depending on whether this object is assumed to have a pure H
($T_{eff}$ $=$ 2977 K) or pure He ($T_{eff}$ $=$ 6494 K) atmosphere.
Neither fit is satisfactory (see Figure \ref{dcdah}\emph{a}), but when
one uses the preliminary trigonometric parallax to constrain the
model, one obtains a mixed H / He atmosphere with $T_{eff}$ $=$ 5078 K
and log N(He)/N(H) $=$ 2.6.  This object is similar to the cool WD,
LHS 3250, except that this object is nearly two magnitudes brighter
and roughly half the distance (17.1 $\pm$ 0.4 pc) from the Sun than
LHS 3250.

\subsection{WD 0121-429}

This object is a moderately cool DA WD that exhibits Zeeman splitting
in the H$\alpha$ and H$\beta$ absorption lines.  Furthermore, the
absorption lines appear quenched (i.e.~not as strong as the model
would predict).  Again using the preliminary trigonometric distance
(18.2 $\pm$ 0.8 pc) to constrain the model, this object is either a
single 0.3M$_{\sun}$ or a pair of unresolved degenerates each at
0.6M$_{\sun}$.  The double degenerate scenario is the more likely
simply because we do not expect to find many low mass WDs due to
galactic age constraints.  This would also help explain the quenched
absorption lines.  If one assumes a double degenerate with one
component being a DAH and the other being a DC, both contributing
equally to the total luminosity, one can reproduce the spectrum fairly
well using $T_{eff}$ from the SED fit as an input.  Figure
\ref{dcdah}\emph{b} displays the best fit and the resultant magnetic
parameter values.  The viewing angle, {\it i} $=$ 65$\deg$, and the
dipole offset, $a_z$ $=$ 0.06 R$_{star}$ are not well constrained, but
what is important is the fit of the central H$\alpha$ line, which
assumes a 50\% dilution factor.  For a complete description of fitting
WDs with strong magnetic fields, see \citet{1992ApJ...400..315B}.  We
have targeted this object to be observed using the Fine Guidance
Sensors (FGS) on the Hubble Space Telescope (HST) in hopes of
resolving this binary and obtaining fractional masses, which, if
successful, would be the first dynamical mass measured for a magnetic
WD.


\acknowledgements 

The Georgia State University team wishes to thank NASA's Space
Interferometry Mission and the National Science Foundation (grant AST
05-07711) and GSU for their continued support.  J. P. S. would like to
thank the Royal Astronomical Society for its financial support to
attend this meeting.  P. B. is a Cottrell Scholar of Research
Corporation and would like to thank the NSERC Canada for its support.
N. C. H. would like to thank colleagues in the Wide Field Astronomy
Unit at Edinburgh for their efforts contributing to the existence of
the SSS; particular thanks go to Mike Read, Sue Tritton, and Harvey
MacGillivray.  We would like to thank Louis Renaud-Desjardins for his
assistance in modeling.  This work has made use of the SIMBAD, VizieR,
and Aladin databases, operated at the CDS in Strasbourg, France.  We
have also used data products from the Two Micron All Sky Survey, which
is a joint project of the University of Massachusetts and the Infrared
Processing and Analysis Center, funded by NASA and NSF.


\end{document}